\documentclass[aps,pre,preprint,groupedaddress]{revtex4-1}

\usepackage{epsfig}
\usepackage{graphicx}
\usepackage{amssymb}
\usepackage{amsmath}
\usepackage{comment}
\usepackage{textcomp} 
\usepackage{subfigure}
\usepackage{color}
\usepackage{natbib}

\begin{document}

\title{ Hydrodynamics of the Developing Region in Hydrophobic Microchannels: A Dissipative Particle Dynamics Study } 


\author{S. Kumar Ranjith}

\author{B. S. V. Patnaik}





\affiliation{Department of Applied Mechanics, Indian Institute of Technology Madras, India}

\author{Srikanth Vedantam}

\affiliation{Department of Engineering Design, Indian Institute of Technology Madras, India}


\date{\today}

\begin{abstract}
Dissipative Particle Dynamics (DPD) is becoming a popular particle based method to study flow through microchannels due to the ease with which the presence of biological cells or DNA chains can be modeled. Many Lab-On-Chip (LOC) devices require the ability to manipulate the transport of cells or DNA chains in the fluid flow. Microchannel surfaces coated with combinations of hydrophilic and hydrophobic materials have been found useful for this purpose.   In this work, we have numerically studied the hydrodynamics of a steady nonuniform developing flow between two infinite parallel plates with hydrophilic and hydrophobic surfaces using DPD for the first time. The hydrophobic and hydrophilic surfaces were modeled using partial-slip and no-slip boundary conditions respectively in the simulations. We also propose a new method to model the inflow and outflow boundaries for the DPD simulations. The simulation results of the developing flow match analytical solutions from continuum theory for no-slip and partial-slip surfaces to good accord. 

The entrance region constitutes a considerable fraction of the channel length in miniaturized devices. Thus it is desirable for the length of the developing region to be short as most microfluidic devices such as cell or DNA separators and mixers are designed for the developed flow field. We studied the effect of a hydrophilic strip near the inlet of a microchannel on the effective developing length. We find that the presence of the hydrophobic strip significantly reduces the developing length.
\end{abstract}

\pacs{}

\maketitle

\section{Introduction}
Miniaturization of fluidic devices from bench-top to palm-top size 
has been progressed considerably in recent years \cite{whitesides2006}. 
The advent of novel micro- and  nano-manufacturing  and fabrication 
techniques has equipped scientists and engineers the ability to 
manipulate the transport of micro/nano liters of fluid through the 
micro/nano channels in these devices \cite{squires2005}. The 
applications of such microfluidic devices are spread through fields 
like electronic-chip cooling, chemical synthesis, targeted cell 
isolation, bio-particle separation processes, chromatography, 
micro-particle sorting, micro-reaction, micro-mixing, 
genomic/proteomic studies and others 
\cite{whitesides2006,sanders2000,dicarlo2009,toner2005,nagrath2007,elali2006}. 
Shrinking the size of table-top labs has led to the evolution of new kinds 
of `on-chip' bio-assays such as lab-on-chip, blood-on-chip, cell-on-chip 
and neurons-on-chip \cite{whitesides2006,toner2005,elali2006,soe2012}. 
The advantages of these microfluidic devices 
are manifold: they are portable, fast, affordable,  accurate and energy 
efficient. The small sample and reagent volumes required and  ease of 
use by non-experts also  equip  them to cater to `point-of-care' needs. 
Fluid flow through channels of a few microns in size  is a common feature of all these devices.

The microchannels are, in many applications, coated with hydrophobic materials \cite{squires2005,rothstein2010}. Hydrophobicity arises when the surface energy of the solid-liquid interface is high and is usually measured from the contact angle made by a sessile drop on a surface \cite{gennes2002,cottin2005}.  Hydrophobic coatings on the walls of 
the microchannels facilitate larger flow rates compared to 
hydrophilic counterparts for the same pressure drop as they 
offer less resistance to flow \cite{lauga2003}. Hydrophobic 
surfaces can also amplify electro-kinetic pumping, aid passive 
chaotic mixing and also mitigate the possibility of choking 
or adhesion of suspended analytes \cite{stroock2002,feui2010,ou2007}. 
On hydrophobic surfaces, the traditional no-slip boundary 
condition is not valid and, instead, the fluid is modeled with a finite 
velocity at the wall. Navier proposed a generalized boundary 
condition to model the velocity of fluid ($u$) tangential to the wall,  
by assuming it to be proportional to the surface shear stress \cite{tretheway2002},
\begin{equation}
u=\beta\left(\frac{\partial u }{ \partial y} \right)_{y=0},
\label{navier}
\end{equation}
 where the {\em slip-length} $\beta$ is the distance from the surface to the point where the linearly extrapolated velocity profile vanishes.  The slip-length can be used to characterize the type of flow in channels; if $\beta=0$ the flow is stick-flow (i.e. no slip), if $\beta=\infty$ the flow is plug-flow (e.g. shear free boundary) and any value of $\beta$ between these two would represent partial slip-flow.  In practical terms, the slip length at hydrophobic surfaces varies from a few nanometers to a few microns \cite{lauga2007}. A slip length of up to 185 $\mu m$  was reported in an experimental  study  \cite{choi2006}, which is comparable to the size of boundary layers in macroscopic regime. In the macroscopic regime the no-slip boundary condition ($i.e.$ velocity of the fluid at solid surface is equal to the wall velocity) captures the physics of flow adequately \cite{granick2003,rothstein2010}. However, for flows in microchannels,  partial-slip boundary conditions have to be applied at hydrophobic surfaces. In the remainder of this paper, we will use hydrophillic to refer to surfaces on which slip length is insignificant and the no-slip boundary condition is appropriate and hydrophobic for surfaces with a finite partial slip.

Numerical simulation of flow through microchannels  is important for understanding the underlying physics of these flows, as well as minimizing effort and expense of experiments, especially during the design and optimization of microfluidic devices.
The range of numerical simulation methods spans continuum based computational fluid dynamics (CFD) to atomistic level molecular dynamics (MD). Modeling flow through microchannels based on the continuum assumption suffers from several drawbacks. First, for flows with high Knudsen number ($Kn$) the continuum approximation may begin to fail \cite{karniadakis2002}. Second, many of the flows include the presence of mesoscale particles such as DNA or individual biological cells.  Treating the interaction of such second phase particles with the fluid medium and including  the Brownian effects due to random thermal fluctuations becomes computationally prohibitive in CFD calculations. Microscopic modeling of above problems using MD is also not a practical choice as it is much too  detailed and computationally expensive. As alternatives, discrete  computational schemes like stochastic rotation dynamics, Brownian dynamics, lattice-Boltzmann method, smoothed particle dynamics, DPD have been developed primarily for the spatio-temporal scales relevant to these situations \cite{mbary2009}.  Among these discrete methods, DPD is a popular mesoscale scheme which bridges the gap between the macroscopic CFD and the microscopic MD \cite{groot1997}.

DPD was introduced to study the dynamics of complex fluids such as colloids, soft matter and polymers \cite{hooger1992}. A cluster of atoms or molecules are considered to form a single particle in the DPD scheme. The positions of the particles are  updated in a Lagrangian framework, and thus DPD can be viewed as a coarse-grained version of MD. In the initial version of the method, the DPD particles were treated as point masses and larger sized particles could be modeled only by binding several of the DPD particles together appropriately. A modified version of DPD was introduced later \cite{wen2008} by treating particles as finite sized and solving for the concomitant rotational degrees of freedom. In this finite-size DPD (FDPD) model, non-central and rotational dissipative forces were considered in addition to the central conservative and non-conservative forces used in conventional DPD. 

The incorporation of appropriate wall boundary conditions for DPD has proved to be a challenge \cite{pivkin2006,haber2006,mbary2009}. Discrete methods show density fluctuations near the walls. While such density fluctuations are realistic at the molecular level, they are considered spurious at the mesoscale continuum level at which DPD purports to model fluids. An instantaneous wall boundary (IWB) model introduced by Ranjith et al.  \cite{ranjith2012} proved useful for modeling no-slip and partial-slip  with minimum near wall fluid property perturbations. Inflow and outflow boundary conditions have also recently been modeled in DPD \cite{lei2011}.  

The hydrodynamics of the developing region inside channels 
with {\em no-slip surfaces} has been studied experimentally \cite{ahmad2010}, 
analytically \cite{sparrow1964} and numerically using CFD \cite{chen1973}. 
Even though there have been considerable efforts to study the fluid 
transport in hydrophobic microchannels \cite{lauga2007,rothstein2010} 
in the fully developed regime, not much attention had been paid to the 
entrance effects with a few exceptions  \cite{bayraktar2006,suman2008}. 
Since the entrance length is proportional to Reynolds number $Re$, it can reasonably be 
ignored in low $Re$ flows. While typical microflows are characterized 
by $Re<30$ \cite{stone2001}, in a few microfluidic  applications like 
micro heat-exchangers, micro-mixers etc the $Re$ reaches the order of 
a few hundreds \cite{stroock2002b,nguyen2005,ou2007}. Wall shear stress effects and velocity distributions 
vary significantly at the entrance and these may eventually  affect the 
separation efficiency of the microfluidic processes \cite{bayraktar2006}. 
Moreover, the entrance region in hydrophobic channels is much 
longer than hydrophilic channels \cite{suman2008}. 
Reduction of entrance length of hydrophobic microchannels is very 
important for the  design of some types of LOC devices. 
In the present work, we study the hydrodynamics of developing flow 
between two parallel plates with hydrophilic and hydrophobic 
surfaces using FDPD method for the first time. We also study 
the effect of hydrophilic patches in the entrance of the channel 
and their effect in reducing the developing length in microchannels.

 In  section \ref{analytical}, analytical solutions for flow 
 between two infinite parallel plates with partial slip are 
 summarized briefly following \cite{sparrow1964,suman2008,duan2010}. 
 In section \ref{scheme}, the governing equations of the FDPD method are 
 summarized. The details for the implementation of slip-wall boundary conditions
 and inflow and outflow boundary conditions are also presented. The FDPD simulation results  
 are then compared with the analytical solution of the developing flow 
 for both no-slip and slip flows in section \ref{entrance}. The velocity profiles and the developing 
 length compare well with the analytical solutions. 
 We then study flow in a channel with hydrophobic walls with a short 
 hydrophilic strip at inlet in section \ref{entrance_strip}. We find 
 that the hydrophilic strip at the entrance shortens the development length significantly.

\section{ Analytical solution from continuum theory }
\label{analytical}

In this section, we briefly summarize the analytical solution of developing flow between parallel flat plates following Sparrow et al \cite{sparrow1964}, Chakraborty and Anand \cite{suman2008},  and Duan  and Muzychka \cite{duan2010} closely.
Consider pressure driven flow between two infinitely long parallel plates separated by $H=2h$. The 
steady, incompressible flow is governed by the mass balance 
\begin{equation}
 \frac{\partial u}{\partial x}+\frac{\partial v}{\partial y}=0,
\end{equation}
and momentum balance 
\begin{equation}
 u\frac{\partial u}{\partial x}+v\frac{\partial v}{\partial y}=-\frac{1}{\rho}\frac{d p}{d x}+\nu \frac{{\partial}^2 u}{{\partial y}^2},
\end{equation}
equations in two dimensions, where $u$ and $v$ are velocities in $x$ and $y$ direction, $\rho$ is the density and $\nu$ is the kinematic viscosity of fluid. The linearized momentum equation \cite{sparrow1964} is of the form 
\begin{equation}
 \nu \frac{{\partial}^2 u}{{\partial y}^2}=\overline{u}\frac{\partial u}{\partial x}+ \frac{\nu}{h}{\left(\frac{\partial u}{\partial y}\right)}_{y=h}
\end{equation} where $\overline{u}$ is the average cross sectional velocity.
The analytical solution of the governing equation  was obtained following Sparrow et al. \cite{sparrow1964} by assuming a no-slip boundary condition. Later, Chakraborty and Anand \cite{suman2008},  and Duan  and Muzychka \cite{duan2010} assumed that the fluid has a finite velocity at the wall  and is modeled by the Navier boundary condition (Eq. \ref{navier}) \cite{suman2008,duan2010}. The slip-length $\beta$  was modeled as a function of  properties of the gas layer adjacent to the wall in case of flow between hydrophobic surface by \cite{suman2008} and as a function of {\it Kn} in case of fluid flow through  microchannels  for $0.001<Kn<0.1$ by \cite{duan2010}.
The governing equations are nondimensionalized using the hydraulic diameter $D_h$ (which is $2H$ for parallel plates), half width $h$ and average velocity $\overline{u}$. Hence the dimensionless parameters are taken as  $\xi=x/\phi$, $\eta=y/h$, $\beta'=\beta/h$ and $U=u/\overline{u}$  where $\phi=D_h/(\rho\overline{u}D_h/\mu)$. The dimensionless form of Navier boundary condition is given by $ U=\beta' \left( {dU}/{d\eta}\right)$. The dimensionless steady velocity profile is a function of both spatial coordinates $\xi$ and $\eta$ and is given by \cite{suman2008}
\begin{equation}
U(\eta,\xi)=\frac{6\beta'}{6\beta'+2}+ \frac{3(1-\eta^2)}{6\beta'+2}+\sum_{i=1}^{\infty} \frac{2[\alpha_i \cos(\alpha_i\eta)-\sin(\alpha_i)]\exp(-16{\alpha_i}^2\xi)}{{\alpha_i}^2 \sin(\alpha_i)[1+3\beta'+{\alpha_i}^2\beta']}.
\label{suman}
\end{equation}
The eigenvalues $\alpha_i$ satisfy
\begin{equation}	
\tan (\alpha_i)=\frac{\alpha_i}{(1+\beta'{\alpha_i}^2)}.
\label{eigen}
\end{equation}
The eigenvalues for partial slip ($\beta'\ne 0$) and no-slip ($\beta' =0$) cases are listed in the appendix.


\section{simulation procedure and boundary conditions}
\label{scheme}
\subsection{Finite-size dissipative particle dynamics}
\label{fdpd}
In this section, the formulation of FDPD model presented by Pan et al. \cite{wen2008} is summarized. 
The domain of interest consists of $N$ DPD particles of finite size with a number density $\rho$.
 In this model, a set  of molecules constitute an FDPD particle having a mass $m_i$ ($i=1,\ldots,N$) and mass moment of inertia  $I_i$ ($i=1,\ldots,N$). The degree of coarse-graining depends on the degree of spatio-temporal detail required. Each FDPD particle obeys Newton's laws of motion and the translational motion is governed by the linear momentum equation
\begin{equation}
m_i \frac{d{\bf v}_i}{dt}={\bf f}_{i},
\label{linear}
\end{equation} 
where ${{\bf v}_i}$ and  ${\bf f}_{i}$ are, respectively, the velocity of and the force on the $i$th particle.
As the particles are of finite size, the angular momentum equation is enforced by
\begin{equation}
I_i \frac{d{\boldsymbol \omega}_i}{dt}=-\sum_{j\neq i} \lambda_{ij}{\bf r}_{ij}\times {\bf f}_{ij},
\label{angular}
\end{equation} 
where ${\boldsymbol \omega}_i$ is the angular velocity and ${\bf f}_{ij}$ is the effective force exerted
on the $i$th particle by the neighboring $j$th particle, at a distance ${\bf r}_{ij}={\bf r}_i-{\bf r}_j$ . The tangential forces are assumed to impart torques on the particles in proportion to the particle radii $R_i$ and thus $\lambda_{ij} =R_i/(R_i+R_j)$.
The position, linear, and angular velocities of each particle is determined by the total force exerted by the surrounding particles within a certain finite cut-off radius $r_c$. In this scheme, the contribution from four types of forces (central ($C$), translational ($T$), rotational ($R$) and stochastic ($S$)) are considered
\begin{equation}
 {\bf f}_{ij}={{\bf f}_{ij}^{C}+{\bf f}_{ij}^{T}+{\bf f}_{ij}^{R}+{\bf f}_{ij}^{S}}.
\end{equation}
The total force on the $i$th particle due to the surrounding particles is given by
\begin{equation}
\tilde{\bf f}_{i}=\sum_{j\ne i}{\bf f}_{ij}.
\end{equation}
The total force on the $i$th particle may include  external forces (${\bf f}^{E}$) from gravitational, magnetic or electro-osmotic forces, if any,  ${\bf f}_{i}=\tilde{\bf f}_{i}+{\bf f}^{E}$.

The central conservative repulsive force acting along the line connecting centers is taken to be
\begin{equation}
{\bf f}_{ij}^{C}=a_{ij}\Gamma(r_{ij}){\bf \hat{e}}_{ij},
\label{cons}
\end{equation}
where $a_{ij}$ is the repulsion parameter, $r_{ij}=\lvert{\bf r}_{ij}\rvert$ and ${\bf \hat{e}}_{ij}={\bf r}_{ij}/r_{ij}$ is a unit vector.  An appropriate weight function $\Gamma(r_{ij})$ is selected such that the conservative force decreases monotonically to 0 at $r_{ij}=r_c$. Most DPD simulations employ the form of $\Gamma(r_{ij})$ given by
\begin{equation}
\Gamma(r_{ij})=\left\{\begin{array}{ll}
1-\frac{r_{ij}}{r_c},& \quad \text{if} \quad  r_{ij} <r_c,\\
0& \quad \text{if} \quad r_{ij}>r_c.
\end{array}\right.
\end{equation}
The translational force is assumed to have central and non-central dissipative components given by
\begin{equation}
{\bf f}_{ij}^{T}=-{\gamma_{ij}}^{C}{\Gamma}^2(r_{ij})({\bf v}_{ij}\cdot{\bf \hat e}_{ij}){\bf \hat e}_{ij}-{\gamma_{ij}}^{S}{\Gamma}^2(r_{ij})[{\bf v}_{ij}-({\bf v}_{ij}\cdot{\bf \hat e}_{ij}){\bf \hat e}_{ij}],
\label{trans}
\end{equation}
where ${\gamma_{ij}}^{C}$ and ${\gamma_{ij}}^{S}$ are the central and shear dissipation coefficients respectively. This frictional force 
attempts to reduce the relative velocity ${\bf v}_{ij}={\bf v}_{i}-{\bf v}_{j}$ between particles in both directions. The rotational dissipative force is taken to be of the form
\begin{equation}
{\bf f}_{ij}^{R}=-{\gamma_{ij}}^{S}{\Gamma}^2(r_{ij})[{\bf r}_{ij} \times (\lambda_{ij}{\boldsymbol \omega}_{i}+\lambda_{ji}{\boldsymbol \omega}_{j})].
\label{rot}
\end{equation}

Finally, a stochastic force is also accounted through the expression given below
\begin{equation}
{\bf f}_{ij}^{S} \Delta t={\Gamma}(r_{ij})[{\sigma_{ij}}^{C} tr[d{\bf W}_{ij}] \frac{1}{\sqrt{d}} {\textbf{1}}+\sqrt{2}{\sigma_{ij}}^{S} d{{\bf W}_{ij}}^A]\cdot {\bf \hat e}_{ij},
\label{rand}
\end{equation}
where $\Delta t$ is the time step, $d=2$ for two-dimensional simulations and $tr[d{\bf W}_{ij}]$ is the trace of symmetric independent Wiener increment matrix $d{\bf W}_{ij}$ while $d{{\bf W}_{ij}}^A$ is its antisymmetric part. According to the fluctuation dissipation theorem, the random and dissipation coefficients are related by ${\sigma_{ij}}^{C}=\sqrt{2k_BT {\gamma_{ij}}^{C}}$ and  ${\sigma_{ij}}^{S}=\sqrt{2k_BT {\gamma_{ij}}^{S}}$. The stochastic forces and the dissipation forces together act to maintain a constant temperature during the simulations.


\subsection{Wall boundary conditions} 
\label{wall}

As mentioned in the Introduction, DPD simulations have shown spurious density fluctuations at walls when enforcing no-slip boundary conditions \cite{pivkin2006,haber2006}. Recently, a new method of enforcing wall boundary conditions in FDPD simulations has shown substantial reduction in the density fluctuations \cite{ranjith2012}. In this method, when a fluid particle is within the range of influence of the wall, the particle interacts with the closest point on the wall as if there were a wall particle for that time step.  The interaction of the wall particle and the fluid particle is separately specified.  This proved to be a simple method to reduce spurious density variations as well as control slip at the wall.
This method, referred to as the instantaneous wall particle boundary (IWB), was shown to be a computationally efficient procedure for modeling impenetrable walls. The slip velocity at wall was tuned by controlling  the lateral dissipative force component between fluid and wall along the direction tangential to the wall. This is achieved by changing the lateral dissipation coefficient $\gamma_{pw}^{S}=\alpha {(1-{r_{pw}}/{r_c)}}^2 {\gamma_{pp}}^{S}$, which acts only within a distance $r_c$  from the wall. Here $r_{pw}$ is the distance between wall and fluid particle. As in the fluid particle-particle interactions, the dissipative and random coefficients are related by ${\sigma_{pw}}^{S}=\sqrt{2k_BT {\gamma_{pw}}^{S}}$. The slip-length increases as the slip modification parameter $\alpha$ is decreased. Thus boundary conditions ranging from no-slip to a large partial slip could be achieved by tuning a slip modification factor $\alpha$ described in that scheme  \cite{ranjith2012}. We note that, some surfaces achieve superhydrophobicity by enhanced roughness which trap air pockets.  In our simulations, we do not model roughness or the second phase fluids at the wall. Instead we specify an effective slip velocity using the Navier condition given by Eq. (\ref{navier}).

\subsection{Determination of slip-length of a hydrophobic surface}
\begin{figure}

\centering

\begin{center}

    {\includegraphics[scale=1]{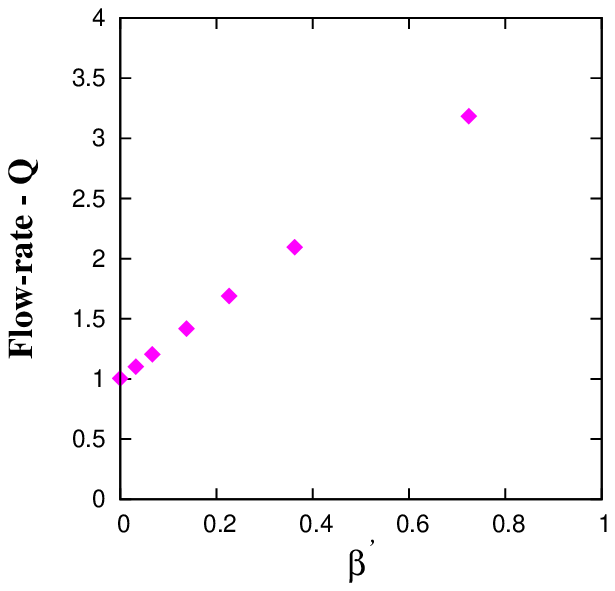}}

\end{center}

\caption{Variation of the flow-rate $Q$ with slip-length $\beta$ for a constant pressure gradient. }
\label{forcebeta}

\end{figure}

A simulation box of $20r_c\times 30r_c$ size taken to be periodic in the stream-wise direction and bound by IWB walls in $y$ direction is used to estimate the slip-length $\beta $. For a constant body force (${\bf f}^E=0.01$), the volume flow-rate per unit area $Q$ is calculated. The $\beta '$ is estimated from the theoretical expression \cite{choi2006b},
 \begin{equation}
 \left(\frac{Q_{slip}}{Q_{no-slip}}\right)_{\Delta P}=1+{3\beta '}
\end{equation}
where slip flow-rate $Q_{slip}$ for different hydrophobic surfaces were obtained by tuning the parameter $\alpha$. 
As seen from Fig. \ref{forcebeta}, the flow-rate increases with increasing $\beta'$. Thus the flow-rate obtained for a certain applied body force increases with increasing partial slip in accordance with the experimental findings reported earlier \cite{choi2006b,lauga2007}.

\begin{figure}

\centering

\begin{center}

    {\includegraphics[scale=1]{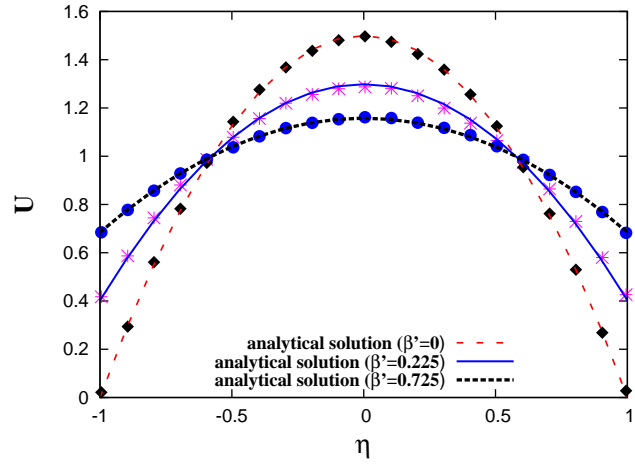}}

\end{center}

\caption{The  fitted analytical solution following \cite{suman2008} (shown with lines)  over FDPD simulated velocity profiles (shown with markers). }
\label{devslip}

\end{figure}

The expression for fully developed velocity profile for slip flow in dimensionless form is given by \cite{suman2008},
\begin{equation}
U(\eta)_{\xi=\infty}=\frac{6\beta'}{6\beta'+2}+ \frac{3(1-\eta^2)}{6\beta'+2}.
\label{developed}
\end{equation} 
 Furthermore, the value of $\beta '$ of a hydrophobic surface in the FDPD simulations is obtained by fitting the analytical velocity profile (Eq. \ref{developed}) to the simulated fully developed velocity profile for slip flow, refer Fig. \ref{devslip}. {The difference in  $\beta '$ determined by both methods is less than $0.66\%$ for the range of slip lengths considered.} For $\alpha=0.15$ the slip-length is found to be  $\beta=2.25r_c$ and corresponding $\beta'= (2.25/10)=0.225$. Unless otherwise specified, this value is used to model all the hydrophobic surfaces mentioned in this work. {The FDPD simulation of two long hydrophilic surfaces separated by $H=30r_c$ and $H=20r_c$ has been carried out to check the effect of the width on the flow-rate. The theoretical scaling of flow-rate per unit area ($Q_{no-slip}^{30r_c}/Q_{no-slip}^{20r_c}$) for $\beta =0$ obtained from the expression \cite{ou2004}
\begin{equation}
 Q_{slip}=\frac{h^2}{3\mu}\left(- \frac{d p}{dx}\right)\left[{1}+\frac{3\beta}{h} \right]
\end{equation} is $2.25$ and that from FDPD simulation is $2.251$. Moreover for slip flow ($\beta =2.25r_c$) the flow-rate ratio from the simulation $Q_{slip}^{30r_c}/Q_{slip}^{20r_c}$ is $1.954$ while theoretical prediction is $1.95$. The FDPD scheme along with the IWB wall is thus able to capture  the developed flow hydrodynamics of hydrophilic and hydrophobic parallel plates.}
 
\subsection{Channel inflow and outflow conditions}
\label{inflow}

In order to compare the FDPD simulations with analytical results, the inflow has been
ensured to be uniform at the inlet with a velocity  ( $\overline{u}=1$) at unit temperature ($k_BT=1$). To maintain the number density of particles in the channel, the particles leaving the channel at the outflow at each time step are reintroduced at the inlet with random positions and random velocities.  The angular and translational velocities are drawn from a uniform distribution in such a way that the system temperature ($k_BT=1$) is not affected by the newly inserted particles. However, it is observed that the reintroduced particles experience forces only from the fluid domain and decelerate. The inlet velocity profiles do not match the analytical results due to this deceleration. In order to overcome this problem, a set of fixed particles (with purely conservative interaction potential) are introduced at the inlet with the same number density of the fluid particles in the domain. These particles provide a balancing force for the particles being reintroduced to the channel and allow the inlet flow to be at the required uniform velocity.

\begin{figure}

\centering

\begin{center}

    {\includegraphics[scale=.6,trim=0mm 0mm 0mm 0mm,clip]{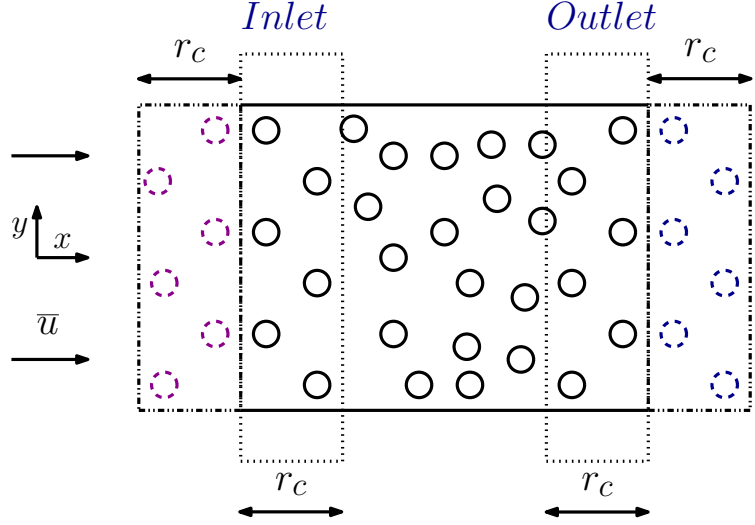}}

\end{center}

\caption{Schematic representation of the inflow and outflow boundaries.}
\label{parallel}

\end{figure}

Similarly, the outflow boundary conditions require balancing forces from outside the fluid domain. In the absence of such balancing forces, the fluid accelerates near the outflow region. To mitigate this effect, particles are fixed outside the outlet of the fluid domain at the same number density as in the fluid domain. These downstream repulsive particles provide the requisite opposing force to the fluid particles at the outlet (schematically shown in Fig. \ref{parallel}).  The repulsive interaction forces exerted by these particles on the fluid particles in outlet is calculated by trial and error. The FDPD velocity profiles were found to match the analytical fully developed profiles to good accord for an inter-particle conservative force parameter value of $a_{po}=\kappa a_{pp}$ with $\kappa=1.2$.

\section{FDPD simulations of developing flow}
\label{entrance}
We devote this section to study the ability of the FDPD model for simulating developing flows in channels with no-slip and partial slip boundary conditions. We consider a steady developing flow between two long parallel plates. The size of the 2D simulation domain is taken to be $20r_c\times400r_c$ and filled with $\rho=3$ particles per unit volume. The effective fluid viscosity is calculated to be $\mu=1$ \cite{ranjith2012}. The short range interactions were calculated with a cut-off radius $r_c=1$.  The Reynolds number $Re=(\rho \overline{u} H/\mu)$ is calculated to be $60$ for the geometry and fluid properties under consideration. The wall and fluid particles are taken to be of the same size and thus the particle size coefficients are $\lambda_{ij}=\lambda_{ji}=\frac{1}{2}$. The maximum repulsion parameter between fluid-fluid $a_{pp}={75k_B T}/{\rho}$ is chosen according to Groot and Warren \cite{groot1997} and fluid-wall $a_{pw}=20$ is taken (`$p$' and  `$w$' represent fluid and wall particles respectively) according to Ranjith {\it et al.} \cite{ranjith2012}. The central and shear coefficients of the dissipative and random forces are taken to be  ${\gamma_{pp}^{C}}={\gamma_{pp}^{S}}={\gamma_{pw}^{C}}=4.5$ and ${\sigma_{pp}^{C}}={\sigma_{pp}^{S}}={\sigma_{pw}^{C}}=3$. The time step was taken to be $dt=0.01$.

All particles in the domain are arranged randomly with zero initial velocity. A body force ${\bf f}^E$ is assumed to act on each particle, which accelerates through the domain. The net momentum ($U$) of domain increases from $0$ to a uniform value of $1$. There domain is decomposed into $400\times 200$ bins in $x$ and $y$ direction across the length and breadth of channel to obtain statistical averages of the velocity inside the domain. The component of velocity in the direction of flow was averaged over $2\times10^5$ iterations to get statistically accurate results.  The force ${\bf f}^E$ is adjusted for each partial-slip boundary condition to maintain the flow rate of $Q=1$. 
For the range of forces applied in the present simulations, a value of $\kappa=1.2$ (ratio of the interaction coefficient of the fluid particle and inflow and outflow boundary particles and inter-particle interaction coefficients) ensured that $Q=1$ and the velocity profiles are very close to the analytical solution. It was observed that for $\kappa<1.15$ the particles close to the outlet accelerate and for $\kappa>1.25$  they decelerate. In both cases the simulated velocity field did not match the analytical solution.

\subsection{Flow in a long hydrophilic channel }
\label{entrance_ns}
A uniform velocity profile at inlet with average velocity $\overline{u}$ at the inlet transforms to a parabolic velocity profile at the outlet with a maximum velocity $1.5 \overline{u}$. Within the developing region the velocity is a function of both $x$ and $y$. When the flow is fully developed, velocity profile is given by Eq. (\ref{developed}) and remains same further downstream.

\begin{figure}

\centering

\begin{center}

  \begin{tabular}{ l  l  l   }
{\bf a}&  & {\bf b} \\ 

    ${\includegraphics[scale=1]{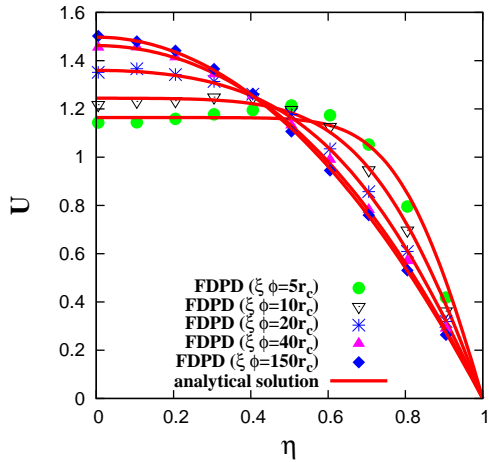}}$ & &${\includegraphics[scale=1]{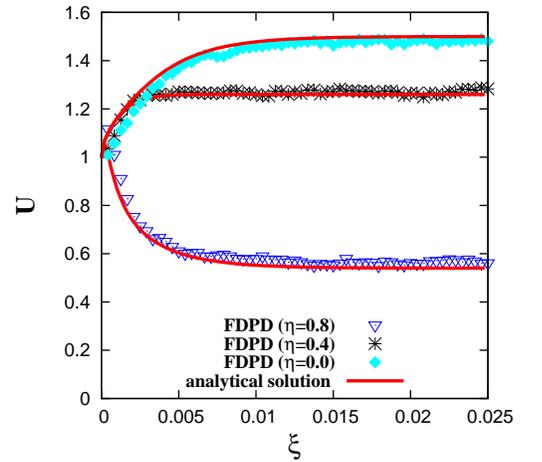}}$    \\


  \end{tabular}

\end{center}

\caption{The velocity profiles in a hydrophilic microchannel ($\beta '=0$) along (a)  span-wise   ($\xi=$ constant) and (b) stream-wise  ($\eta=$ constant) directions.}
\label{hydro}

\end{figure}

The no-slip condition at the wall is obtained by modeling the solid boundary with a slip modification factor $\alpha=3$ as reported in our earlier work \cite{ranjith2012}. The inflow and outflow boundary conditions were implemented as discussed in section \ref{inflow}. The velocity profiles at different axial positions ($\xi=$ constant) are plotted in Fig. \ref{hydro} (a), along with the analytical solution. In Fig. \ref{hydro} (b), the velocity profiles  at different heights ($\eta=$ constant) are extracted and compared with the analytical solution. The FDPD simulation and analytical solutions were found to be in good agreement. The channel length was chosen  to be $L_{c}=20H$ to minimize end effects. Some end effects are apparent within $H/4$ from the outlet in the form of velocity fluctuations (less than $12\%$ of the maximum velocity).  

 \subsection{Flow in a long  hydrophobic channel }
\label{entrance_s}
The effective slip at the hydrophobic surface is obtained by choosing an appropriate parameter $\alpha$ as discussed in section \ref{wall}. Due to the non-zero velocity at the wall, the acceleration of the central laminar core is less compared to that of the no-slip case. The velocity gradients produced by partial-slip wall are smaller than the no-slip walls. Due to this, the developing length $L_e$ for superhydrophobic microchannels is greater than in hydrophilic channels \cite{suman2008}.

\begin{figure}

\centering

\begin{center}

  \begin{tabular}{ l  l  l   }

{\bf a}&  & {\bf b} \\ 
    ${\includegraphics[scale=1]{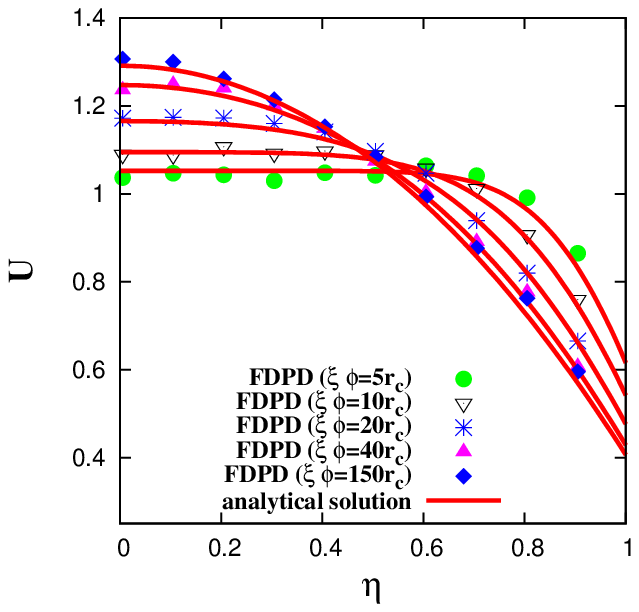}}$ & &${\includegraphics[scale=1]{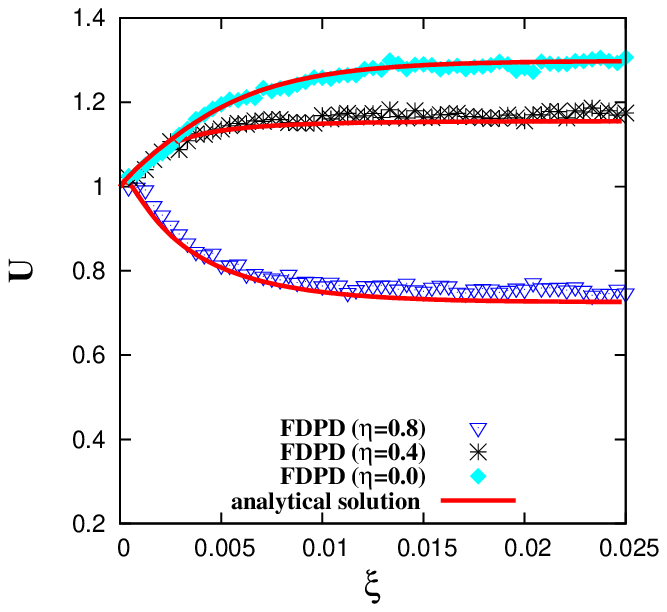}}$    \\


  \end{tabular}

\end{center}

\caption{The velocity profiles in a  hydrophobic microchannel ($\beta '=0.225$) along (a)    ($\xi=$ constant) and (b) stream-wise  ($\eta=$ constant) directions.}
\label{superhydro}

\end{figure}

The velocity profiles obtained through the FDPD simulation is compared to the analytical solution given by Eq. (\ref{suman}). The  theoretical and computational results are in good agreement as can be seen in Fig. \ref{superhydro}.  This simulation shows that the FDPD scheme, in combination with IWB wall, can capture the hydrodynamics of the steady non-uniform developing region of partial-slip flow accurately. The effect of a no-slip region at the entrance on the hydrodynamics of fluid flow  between two hydrophobic surfaces is discussed in the next section.

\section{Entrance region with a hydrophilic strip}
\label{entrance_strip}

The developing length or entrance length is the stream-wise distance from the inlet to the point 
at which the boundary layers formed on both the walls merge. In an engineering sense,
this is quantified by the distance along the flow direction at which the centerline velocity reaches 99\% of the maximum fully developed velocity. 
There are several empirical correlations for the entrance length as a function of the Reynolds 
number for no-slip channels \cite{chen1973,duan2010}.
 For moderate Reynolds numbers, the developing length constitutes a considerable portion of a 
miniaturized LOC device, and hence a reduction in the entrance length of the 
microchannel is highly desirable. For partial-slip flow, the development of the 
flow field occurs over a much greater length compared to no-slip wall boundary 
conditions. This is because the acceleration of the central core for a no-slip wall 
is greater than that of a slip-wall for the same $Re$. 
The greater shear stress at the wall for a hydrophilic surface enables the presence of 
the wall to be felt inside the domain over a shorter distance 
compared to their superhydrophobic counterparts. The latter have a smaller magnitude of the
velocity gradient near the walls.  We explore the hydrodynamics of flow 
between two long parallel superhydrophobic surfaces with a 
hydrophilic strip at the inlet in this section.

Experiments on the entrance hydrodynamics in microchannels with no-slip boundary conditions with aspect 
ratio $H/W=1$ ($W$ being the width of the channel) were recently  reported by Ahmad and Hassan \cite{ahmad2010} for a hydraulic radius ($D_h=H$)  ranging from $100\mu m$ to $500\mu m$  over a range of $Re$ numbers from  $0.5$ to $ 200$. 
The entrance length of a  hydrophilic microchannel with 
$D_h=200\mu m$ and $Re=60$ is interpolated from the 
empirical correlation obtained from their experimental 
data (Eq. (6) of Ref. \cite{ahmad2010})  is 943 $\mu m$. 
Using an analytical approach, Chakraborty and Anand \cite{suman2008} 
have presented a correlation that relates the developing 
length and Reynolds number of slip surfaces,
\begin{equation}
 \frac{L_e}{D_h}=\frac{0.63}{0.035Re+1}+0.044Re(1+1.675\beta '+2.3125{\beta '}^2).
\label{le}
\end{equation}
Here, $D_h=H$, for a given slip length ($\beta '$).   However, to the best of our knowledge there is no experimental data available till date on the hydrodynamics of entrance region of 
microchannels with hydrophobic surfaces. Hence, the above
analytical solution in Eqn.\ref{le} is used to determine
the developing length under partial-slip conditions. The ratio of the hydrodynamic 
development length for hydrophobic (partial-slip) to hydrophilic (no-slip) surfaces is calculated to be
\begin{equation*}
 {\left(\frac{L_e^{\beta '=0.225}}{L_e^{\beta '=0}}\right)}_{Re=60}=1.46.
\end{equation*}  
Thus the developing lengths 
of hydrophobic channel is about 50\% larger than hydrophilic channel for the same $Re$. It was reported 
in Ref. \cite{choi2006} that, a nano-turf created by coating a surface with 
Teflon can produce a hydrophobic surface with  
$\beta\approx20\ \mu m$. For such a hydrophobic material 
with $\beta '=0.225$ ($\beta=22.5\ \mu m$, $h=D_h/2=100\ \mu m$), 
and a typical $Re=60$ would have a  $L_e\approx 1.46\times943 \ 
\mu m=1377\ \mu m$. For an `on-chip' device which is already 
small in size, this undesirable entrance region would constitute 
a major portion of the channel. This effect increases as the number 
of parallel microchannels (which is typical for most LOC devices)  increases to achieve large throughput.

\begin{figure}

\centering

\begin{center}

    {\includegraphics[scale=.8]{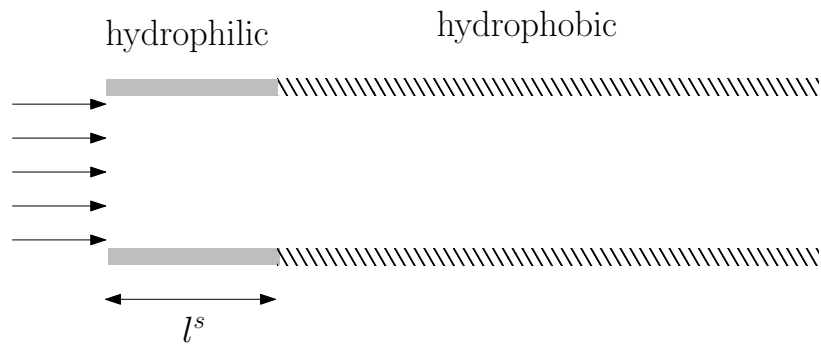}}

\end{center}

\caption{Schematic sketch of a  hydrophobic channel with a hydrophilic strip of length  $l^s$ at the inlet.}
\label{sketchns}

\end{figure}

 In this section the effect of introduction of a no-slip strip having length $l^{s}$ at the inlet, just before the hydrophobic surface (schematically shown in Fig. \ref{sketchns}) is discussed. The velocity profiles obtained through the DPD simulations closely follows the analytical solution, so Eq. (\ref{le}) was used to estimate $L_e$ for slip flow. Thus for a hydrophobic surface with effective slip length $\beta '=0.225$ and $Re=60$ has a developing length of $83r_c$. Fixing a hydrophilic strip accelerates the central core faster compared with hydrophobic wall as large shear gradients are formed near no-slip surfaces.
\begin{table}
 \caption{The entrance length for mixed hydrophobic-hydrophilic channels $L_e^{s}$ with different lengths of hydrophilic strips $l^{s}$ at the inlet.}
\begin{tabular}{|c|c|c|}
\hline 
Sl. No. & $l^{s} $ & $L_e^{s}$\\ \hline
1	& $0r_c$ & $83r_c$ \\ \hline
2	& $5r_c$ & $50r_c$ \\ \hline
3	& $10r_c$ & $30r_c$ \\ \hline
4	& $15r_c$ & $55r_c$ \\ \hline
5	& $20r_c$ & $65r_c$ \\ \hline
6	& $25r_c$ & $100r_c$ \\ \hline
\end{tabular}
\label{dle}
\end{table}

\begin{figure}

\centering

\begin{center}

    {\includegraphics[scale=1.1,trim=13mm 0mm 10mm 2mm,clip]{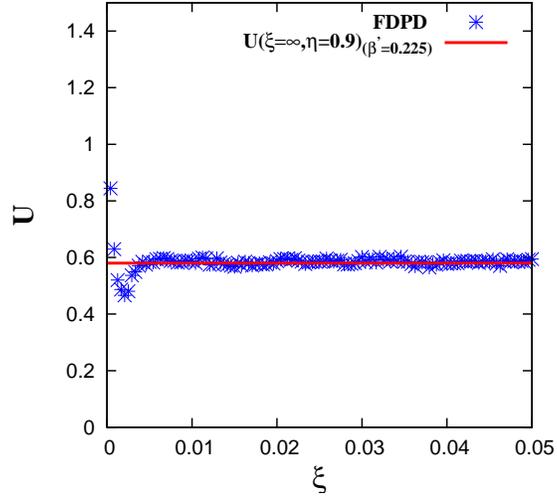}}

\end{center}

\caption{Comparison of the simulated velocity profile of hydrophobic channel with hydrophilic inlet strip of length $l^{s}=10r_c$ with the fully developed theoretical  velocity profile at $\eta=0.9$ from Eq.~(\ref{developed}).}
\label{noslip_strip}

\end{figure}

\begin{figure}

\centering

\begin{center}

  \begin{tabular}{  l r    }
{\includegraphics[width=8cm,height=3.6cm]{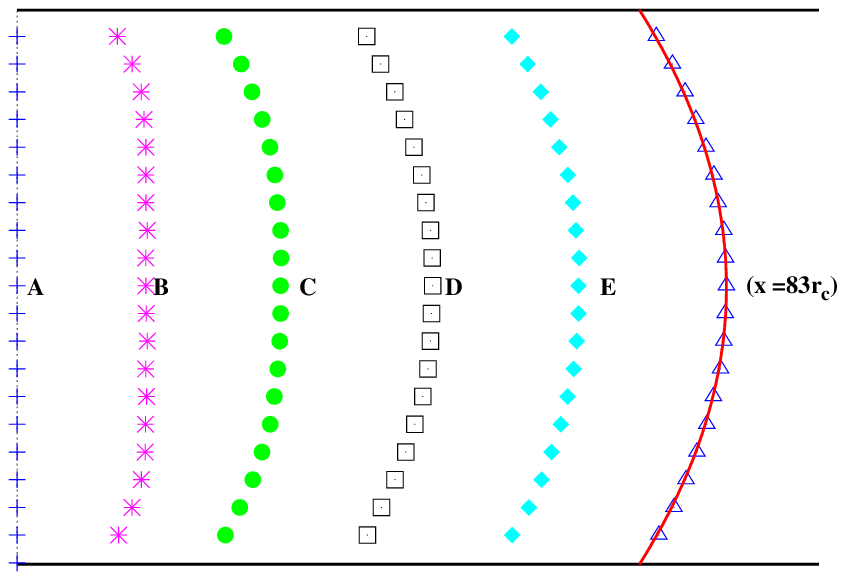}} &{\bf (a)} \\
{\includegraphics[width=8cm,height=3.6cm]{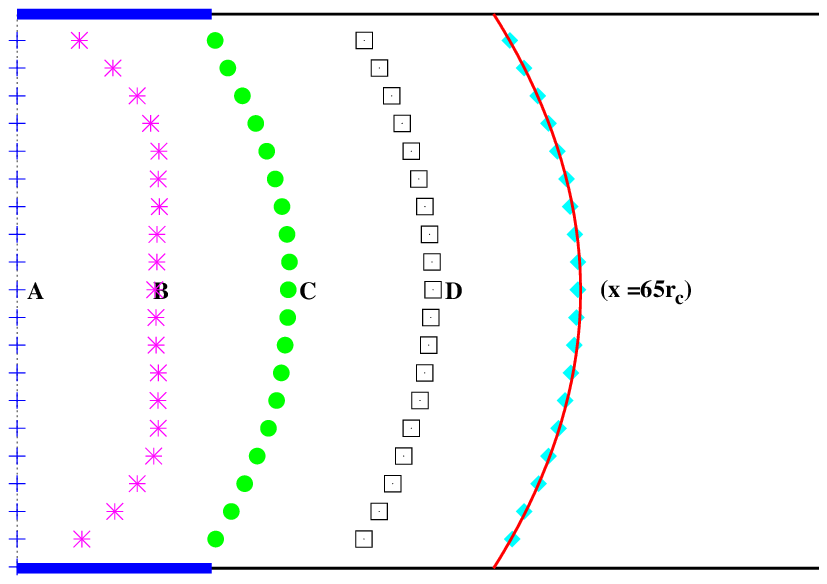}}&{\bf (b)} \\
{\includegraphics[width=8cm,height=3.6cm]{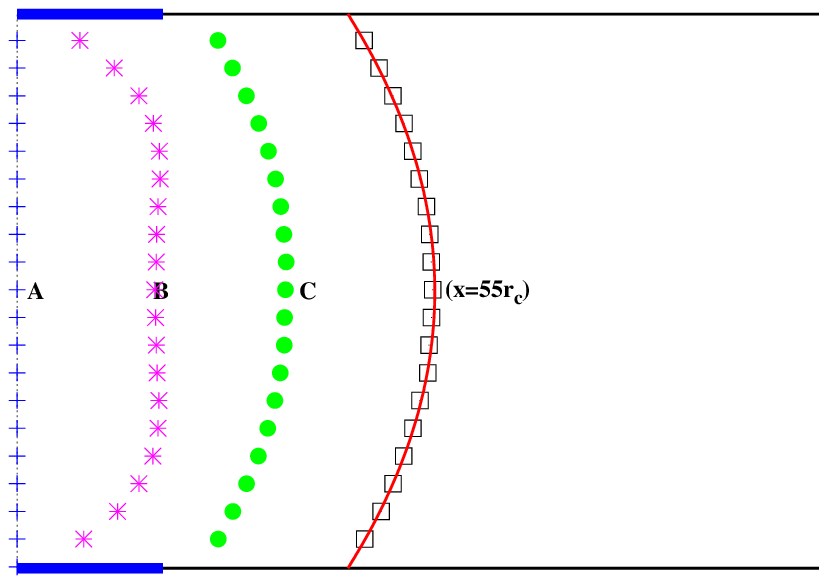}}&{\bf (c)} \\
{\includegraphics[width=8cm,height=3.6cm]{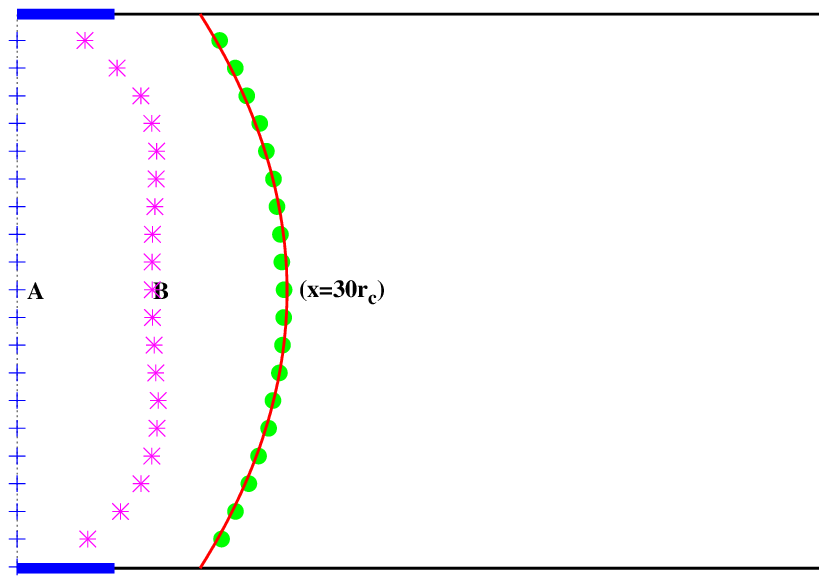}}&{\bf (d)} \\
{\includegraphics[width=8cm,height=3.6cm]{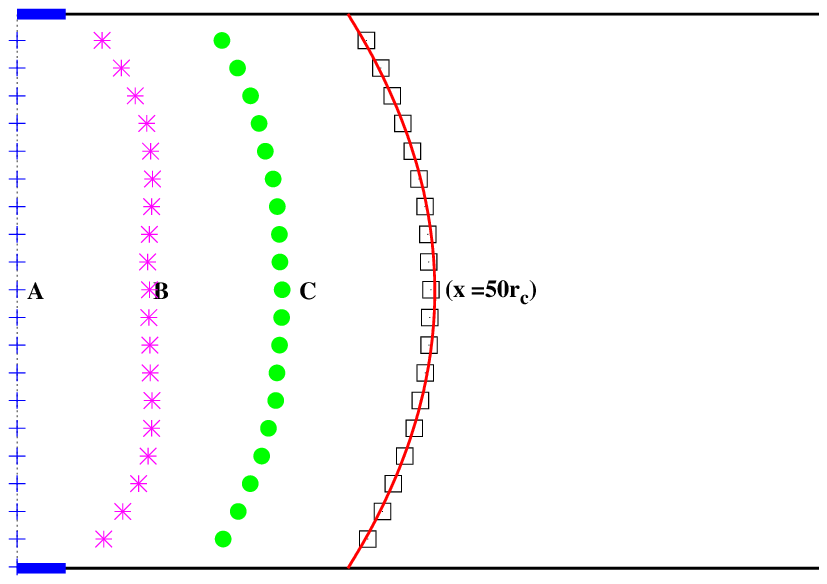}}&{\bf (e)} \\
  \end{tabular}

\end{center}

\caption{Comparison of the simulated velocity profiles (shown with markers) at different locations from the inlet in a mixed hydrophobic channel for various lengths of hydrophilic strips at the inlet: (a) $l^{s}=0r_c$, (b) $l^{s}=20r_c$, (c) $l^{s}=15r_c$, (d) $l^{s}=10r_c$, (e) $l^{s}=5r_c$. The profiles at $x=0$ are marked A, $x=10r_c$ by B, $x=30r_c$ by C, $x=55r_c$ by D and $x=65r_c$ by E in the figures. The fully developed analytical velocity profile for $\beta = 2.25r_c$ is shown in each case using a solid line for comparison.}
\label{lecompare}

\end{figure}

To study the influence of the no-slip strips, a number of DPD simulations were 
carried out by varying the initial hydrophilic strip length  from $5r_c$ to $25r_c$ at $Re=60$. 
As the stream-wise velocity gradient is maximum near the wall, the entrance length 
of a mixed hydrophilic-hydrophobic surface was found by monitoring the velocity near
the wall. The transition between the hydrophilic and hydrophobic surface results in velocity
fluctuations.  These fluctuations are minimum along the centerline ($\eta=0$)
and greatest at the wall. So the developing length with a
hydrophilic strip ($L_e^{s}$) at the entrance, is estimated by 
determining the distance from the inlet to the point where the axial velocity for 
$\eta=0.9$ becomes constant. The effect of the length of the inlet hydrophilic strip on the developing length is given in Table \ref{dle}. It was found 
that for a $10r_c$ hydrophilic strip, the developing length
reduces  from $83r_c$ to $30r_c$ (see Fig. \ref{noslip_strip}) and the percentage 
reduction in the developing length is $\left(\frac{L_e-L_e^{s}}{L_e}\right)\approx 66\%$,
although the  portion of hydrophobic surface replaced by hydrophilic surface 
is only ($l^{s}/L_e)\approx\frac{1}{8}\approx 12.5\%$.  
Thus, the combination of hydrophilic-hydrophobic surfaces drastically 
reduced the developing length $L_e$. The full development of velocity 
profile of such an arrangement takes place over a shorter distance than that 
of pure hydrophobic surfaces, as shown in Fig. \ref{lecompare}.  This finding is expected to be beneficial for the 
design of microfluidic devices and to optimize the size of 
LOC devices with hydrophobic channels. {It is noteworthy that, from the analytical solution (Eq. \ref{suman}) the central line velocity at the end of hydrophilic strip ($x=10r_c$) is $1.244$ and the fully developed centerline velocity of hydrophobic surface is $1.3$. We infer that, the $l^s$ should be long enough to accelerate the central core of hydrophilic region nearer to the developed velocities of the hydrophobic surfaces. If the $l^s$ is too long the center region accelerates, and take a longer distance to decelerate and reach a developed state as shown in Fig. \ref{lecompare}(b). Conversely, if the $l^s$ is short the acceleration of central core is not enough to reach the uniform fully developed state and may take a longer distance, as seen in Fig. \ref{lecompare}(e).}

\section{Summary and Conclusions}
In this work, a modified DPD method has been shown to effectively capture the hydrodynamics of developing flows in microchannels with no-slip and partial-slip boundaries. The simulations with no-slip and partial slip wall 
boundary conditions  were shown to have excellent agreement 
with analytical results. A new method to model inflow boundary condition is 
proposed to obtain a uniform inlet velocity profile. Similarly, the outflow 
boundary conditions were modified to prevent the fluid from accelerating 
out of the domain. The presence of a small hydrophilic strip at the 
inlet of a hydrophobic microchannel was  found to 
significantly reduce the development length of the remaining hydrophobic 
channel. This finding can potentially be used by the designers of LOC 
devices to optimize the size of the microfluidic devices.


%
%


%

\clearpage
\appendix{APPENDIX}

\begin{table}[h!b!p!]
\caption{The eigenvalues obtained from Eq. \ref{eigen} used to calculate the analytical solution of velocity.}
\begin{center}
  
\begin{tabular}{|c|c|c|}
\hline 
$i$   & $\alpha_{i}^{\beta=0}$  &$\alpha_i^ {\beta '=0.225}$  \\ \hline
1	& $4.4934$ & $3.8666$ \\ \hline
2	& $7.7253$ & $6.8198$ \\ \hline
3	& $10.9041$ & $9.8327$ \\ \hline
4	& $14.0662$ & $12.8903$ \\ \hline
5	& $17.2208$ & $15.9749$ \\ \hline
6	& $20.3713$ & $19.0758$ \\ \hline
7	& $23.5195$ & $22.1871$ \\ \hline
8	& $26.6661$ & $25.3054$ \\ \hline
9	& $29.8116$ & $28.4286$ \\ \hline
10	& $32.9564$ & $31.5552 $ \\ \hline
11	& $36.1006$ & $34.6845$ \\ \hline
12	& $39.2444$ & $37.8157$ \\ \hline
13	& $42.3879$ & $40.9485$ \\ \hline 
14	& $45.5311$ & $44.0826$ \\ \hline 
15	& $48.6741$ & $47.2176$ \\ \hline 
16	& $51.8170$ & $50.3534$ \\ \hline
17	& $54.9597$ & $53.4898$ \\ \hline
18	& $58.1023$ & $56.6269$ \\ \hline
19	& $61.2447$ & $59.7644$ \\ \hline
20	& $64.3871$ & $62.9023$ \\ \hline
21	& $67.5294$ & $66.0406$ \\ \hline
22	& $70.6717$ & $69.1791$ \\ \hline
23	& $73.8139$ & $72.3180$ \\ \hline
24	& $76.9560$ & $75.4570$ \\ \hline
25	& $80.0981$ & $78.5963$ \\ \hline
\end{tabular}
\label{alpha}

\end{center}
\end{table}

\begin{acknowledgments}
SKR gratefully acknowledges the research sponsorship under AICTE-QIP (Government of India) scheme.  We thank the developer communities of the following free software: GNU/Linux (Ubuntu), Gfortran and Gnuplot for providing excellent platforms for our computational requirements.
\end{acknowledgments}

\bibliography{refpre}

\listoffigures
\newpage


\end{document}